\documentclass[sigconf]{acmart}

\usepackage{booktabs} 
\usepackage[listings]{tcolorbox}
\usepackage{xcolor}
\usepackage{multirow}
\usepackage{hyperref}
\usepackage{url}
\usepackage{comment}
\usepackage{todonotes}
\usepackage{color}
\usepackage{balance}

\copyrightyear{2022}
\acmYear{2022}
\setcopyright{acmcopyright}\acmConference[SERP4IoT'22]{4th International Workshop on Software Engineering Research and Practice for the IoT}{May 19, 2022}{Pittsburgh, PA, USA}
\acmBooktitle{4th International Workshop on Software Engineering Research and Practice for the IoT (SERP4IoT'22), May 19, 2022, Pittsburgh, PA, USA}
\acmPrice{15.00}
\acmDOI{10.1145/3528227.3528569}
\acmISBN{978-1-4503-9332-4/22/05}

\begin{document}

\title[SE approaches for TinyML based IoT Embedded Vision]{Software Engineering Approaches for TinyML based IoT Embedded Vision: A Systematic Literature Review}

\author{Shashank Bangalore Lakshman}
    \affiliation{%
        \institution{Boise State University}
        \city{Boise}
        \state{ID}
        \country{USA}
}
\email{shashankbangalor@u.boisestate.edu}

\author{Nasir U. Eisty}
\affiliation{%
	   \institution{Boise State University}
	   \city{Boise}
	   \state{ID}
	   \country{USA}
}
\email{nasireisty@boisestate.edu}

\renewcommand{\shortauthors}{S. Bangalore Lakshman et al.}

\begin{abstract}
Internet of Things (IoT) has catapulted human ability to control our environments through ubiquitous sensing, communication, computation, and actuation.
Over the past few years, IoT has joined forces with Machine Learning (ML) to embed deep intelligence at the far edge. 
TinyML (Tiny Machine Learning) has enabled the deployment of ML models for embedded vision on extremely lean edge hardware, bringing the power of IoT and ML together. 
However, TinyML powered embedded vision applications are still in a nascent stage, and they are just starting to scale to widespread real-world IoT deployment. 
To harness the true potential of IoT and ML, it is necessary to provide product developers with robust, easy-to-use software engineering (SE) frameworks and best practices that are customized for the unique challenges faced in TinyML engineering. 
Through this systematic literature review, we aggregated the key challenges reported by TinyML developers and identified state-of-art SE approaches in large-scale Computer Vision, Machine Learning, and Embedded Systems that can help address key challenges in TinyML based IoT embedded vision. 
In summary, our study draws synergies between SE expertise that embedded systems developers and ML developers have independently developed to help address the unique challenges in the engineering of TinyML based IoT embedded vision.

\end{abstract}

%
%



\maketitle

\section{Introduction}
\label{sec:Introduction}
\begin{figure}[htp]
    \centering
    \includegraphics[width=\linewidth]{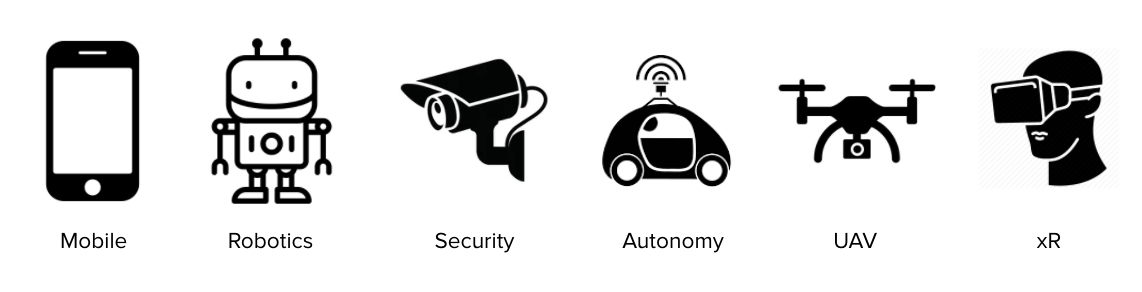}
    \caption{Emerging embedded vision IoT applications}
    \label{fig:apps}
\end{figure}

The desire for a better lifestyle and the growing abundance of challenges faced on the planet have necessitated technological breakthroughs. 
‘Edge Intelligence’ aims to harness the combined power of IoT and ML to solve unique problems at scale, right at the place where the data gets generated \cite{7488250}. 
Millions of ubiquitous smart embedded vision systems (cameras with local computing power and network connectivity) are deployed and will be deployed at the edge in various domestic, industrial, and commercial applications (some examples are illustrated in Figure \ref{fig:apps}). 
IoT sensor networks collect gigabytes of data every minute, and their computational systems must process the data streams in real-time to provide valuable, actionable insights to people, and other systems \cite{7488250}. 
IoT embedded vision systems are being deployed at the far edge for valuable gains in speed, power efficiency, cost efficiency, privacy, and autonomy \cite{7488250}. 
Many previously intractable problems that previously required human experts and sophisticated hardware-software systems for decision-making are now starting to be automated by deploying start-of-art ML models on lean embedded IoT.

SE for ML has evolved over the last decade \cite{giray2021software}, aimed at fielding a plethora of challenges faced by ML developers to harness the power of Deep Neural Networks (DNN). 
Similarly, SE for IoT embedded systems has also evolved over the past decade into a successful research domain \cite{oshana_kraeling_2019}. 
Because TinyML combines ML for computer vision (CV) with IoT embedded systems, engineering, deployment, and maintenance of TinyML applications, requires a well-defined and customized SE approach to address the unique challenges in the domain. 
Without a guiding set of SE approaches for TinyML based IoT embedded vision, product development lifecycle can involve high costs, low productivity, and weaknesses in the field \cite{giray2021software}.
The need for low latency, low power, low cost, and robustness in these applications requires SE processes to be aware of systems architecture, algorithms, and challenges in real-world deployments.
Our research summarizes state-of-art SE approaches applicable to solve the unique challenges in TinyML engineering and proposes a streamlined SE workflow to simplify the development, deployment, and maintenance of IoT embedded vision applications, in addition to suggestions for developer tools.

\section{Background}
\label{sec:Background}
It is projected that a total of 2.5 billion edge AI devices will ship with a TinyML chipset in 2030 \cite{abiresearch}.
Early Proof-of-Concept (POC) applications using TinyML based IoT embedded vision have showcased huge potential to enable smart manufacturing, smart home, smart city, smart devices, infotainment, autonomy, and smart agriculture applications on extremely lean IoT systems. 
However, this approach is yet to scale into massively deployed products and solutions in the real world, as these systems are challenged by portability, robustness, reliability, development costs, and skill gaps \cite{soro2021tinyml}.

The rapid growth in CV applications powered by Deep Learning (DL) approaches is due to DNN-based approaches extracting features automatically from training data, without the need for heuristic rules \cite{lecun_bengio_hinton_2015}. 
Typical CV applications are driven by Convolutional Neural Networks (CNN) as visual information has rich spatial information. 
Trained models also must be trained with the appropriate quantity and quality of data based on an understanding of input data expected in real-world deployments. 
As DL became popular over the decade, there was an uncontrolled growth in the size of DL models.
This growth resulted in steep demand for both computing power and memory availability in training as well as inference systems \cite{szehw4ml}.

Migrating such monstrous DL models into leaner IoT embedded systems for inferencing tasks is a unique challenge that requires a systematic overhaul towards hardware-aware engineering \cite{szehw4ml}. 
Typical advanced MCUs are those that are capable of running highly-optimized, compressed DNN models by leveraging only a few 100 kBs of memory, computing speed in the order of MIPS (Million Instructions Per Second) or GIPS (Giga Instructions Per Second) at a power consumption of <1W \cite{david2021tensorflow}.
`Model reduction', `parameter quantization', 'knowledge distillation', and 'Neural architecture search' are the key tools for overcoming machine learning challenges on the edge \cite{DBLP:journals/corr/abs-2003-11066} \cite{soro2021tinyml} \cite{szehw4ml}.

\begin{figure}[htp]
    \centering
    \includegraphics[width=\linewidth]{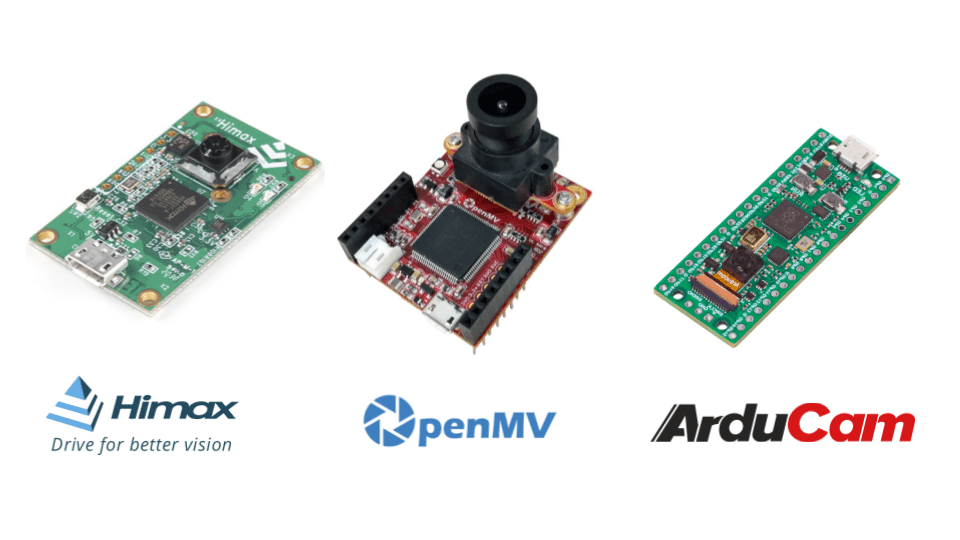}
    \caption{Example IoT embedded vision devices \cite{chaudhary_2021}}
    \label{fig:devices}
\end{figure}

IoT embedded vision applications offer the advantage of low latency, real-time actionable insights, lower cost of data transmission, higher reliability under economy of scale.
TinyML can be broadly defined as ML approaches capable of performing on-device analytics for a spectrum of sensing modalities at “mW” (or below) power range, on lean and battery-operated devices.
A few examples of off-the-shelf IoT embedded vision hardware capable of running TinyML applications are showcased in Figure \ref{fig:devices}. 
A comparison of their key technical specifications is presented in Table \ref{tab:compare}. 
If the TinyML engineering process can be made more efficient, then the lifetime cost of applications can be driven further low by the massive market potential for IoT embedded vision applications.

\begin{table*}[htbp]
    \centering
    \caption{Comparison of hardware specifications of off-the-shelf IoT embedded vision hardware \cite{chaudhary_2021}}
    \begin{tabular}{p{0.1\linewidth} | p{0.28\linewidth} | p{0.28\linewidth} | p{0.28\linewidth}}
        \hline
        IoT device & HIMAX WE-I Plus EVB & OPENMV Cam H7 R2 & ARDUCAM Pico4ML \\
        \hline
        
        Compute & WE-I Plus ASIC (HX6537-A), ARC 32-bit EM9D DSP with FPU @400MHz & ARM 32-bit Cortex-M7 CPU w/ Double Precision FPU @480MHz & RP2040 Dual-core Arm Cortex-M0+ processor @133MHz \\
        \hline
        
        Memory & 2MB flash and 2MB SRAM & 2MB flash and 1MB SRAM & 2MB flash and 256kB SRAM \\
        \hline
        
        Vision sensor & Himax HM0360 AoS TM ultra-low power VGA CCM, 1/6", 640×480 pixels @60FPS & MT9M114: CMOS Image Sensor, 1/6", 640×480 pixels @40FPS & Himax HM01B0 CCM, 1/11", 320 x 240 pixels @60FPS \\
        \hline
        
        Device cost & \$65 & \$65 & \$50 \\
        \hline
        
    \end{tabular}
    \label{tab:compare}
\end{table*}   

\begin{table*}[htbp]
    \centering
    \caption{Key challenges in TinyML based IoT embedded vision engineering}
    \begin{tabular}{p{0.20\linewidth} | p{0.76\linewidth}}
        \hline
        CHALLENGE & DESCRIPTION \\
        \hline
        
        Requirements specifications and design & System design choices: No clear decision framework or benchmark that can guide TinyML based IoT embedded vision application developers with regard to hardware/model design choices to be made for a given application/market/customer requirement \cite{mukherjee_verma_gopani_choudhary_2020}. \\
        
        & Compiler choices: Embracing sophisticated compilers can help optimize for specific MCU targets. However, this affects portability, and hence, it challenges large-scale deployment under availability constraints. Need to enable platform specific optimizations without need for specialized compiler \cite{david2021tensorflow}. \\
        
        & Ensuring the security of embedded devices is a challenge for IoT applications \cite{loukides_2019} \cite{8979377}. \\
        
        & Requirements in the field may evolve, and the system needs to grow its intelligence to meet the goals \cite{giray2021software}. \\
        
        & IoT embedded vision devices need to be "less chatty" to reduce the power consumed by communications~\cite{loukides_2019}. \\
        
        & IoT embedded devices have a high degree of heterogeneity \cite{banbury2021benchmarking}. \\
        
        & Although hand coding and code generation can leverage specific optimizations at the cost of flexibility \cite{banbury2021benchmarking}. \\
        
        & On-device training is challenging in small memory footprint devices \cite{DBLP:journals/corr/abs-2007-11622} \\
        \hline
        
        Data and model search & Lack of curated datasets derived from IoT embedded vision sensors for different applications \cite{banbury2021benchmarking}. \\
        
        & Lack of standard tools to compare performances between different algorithms against different MCU systems \cite{banbury2021benchmarking}. \\
        \hline
        
        Model development and validation & Challenging to assess the impact of different levels of quantization and precision as well as model peak memory requirements \cite{banbury2021benchmarking}. \\
        
        & Due to the wide spectrum of MCU designs, it is beneficial to take a framework with generic compiler features for design portability. However, this generic compiler might not be able to fully utilize unique hardware features offered in some families of MCUs \cite{david2021tensorflow}. \\
        
        & TinyML workflow is not hardware-aware due to limitations in generic frameworks \cite{david2021tensorflow}. \\
        
        & Naive joint optimization techniques such as Neural Architecture Search (NAS), Quantization and Pruning have cross-interactions which may result in sub-optimal results \cite{ba43358608fe4cfcb202718ceee0805c}. \\
        
        & Lack of emulation tools for TinyML engineering increases the time and effort spent on model development~\cite{gielda_2022}. \\
        \hline
        
        Application development and deployment & Application portability across different devices and different vendors is a challenge in TinyML engineering~\cite{reddi2021widening}. \\
        
        & Power profiling is complex since data paths and pre-processing steps can vary significantly between devices~\cite{banbury2021benchmarking}. \\
        
        & TinyML powered applications are lean, and they lack tools for measurement and visualization of data. This imposes additional challenges to debuggability during model development and deployment due to memory constraints \cite{banbury2021benchmarking}. \\
        
        & TinyML capable devices have drastically different power consumption, which makes it harder to maintain accuracy consistently across the devices \cite{banbury2021benchmarking}. \\
        
        & Real-world conditions can rapidly evolve due to changes in the sensing environment or the aging of vision sensors. TinyML improves a significant challenge to online learning due to the low amount of compute available \cite{8979377}. \\

        & Since edge CV applications may be used to drive time-critical decisions, the reliability of TinyML systems needs to be high. Robustness needs to be built into the TinyML models to enable safe fail-over mechanisms in the case of ambiguous environments \cite{8979377}. \\
        
        & Models deployed on TinyML vision systems must be capable of being upgraded without significant intervention or system downtime. \\
        \hline
        
        SE methodologies & Not much is known about efficient, best practices for TinyML application IoT embedded vision product development, deployment and maintenance. \\
        
        &  Agile development practices for TinyML engineering are not well established. \\
        
        & TinyML CV application requires expertise in embedded systems, computer vision, and machine learning. Since TinyML is still in a nascent stage, there is a lack of skilled engineers who have sufficient expertise to address challenges in TinyML engineering for CV apps \cite{reddi2021widening} \\
        \hline
        
        \hline
        
    \end{tabular}
    \label{tab:challengestinyml}
\end{table*}

ML has gained massive popularity with application developers due to the availability of hardware-agnostic open-source ML frameworks such as TensorFlow and PyTorch.
Similarly, TinyML has embraced a framework approach to simplify developer experience and amplify productivity. 
TinyML has multiple frameworks such as TensorFlow Lite Micro (TFLM), uTensor, STM32Cube.AI, and Embedded Learning Library (ELL). 
However, our research only focused on TFLM framework due to its popularity and being independent on the hardware. 

TFLM is an open-source ML inference framework (developed by Google and Harvard University) for running deep-learning models on MCU based embedded systems with very lean computing and memory infrastructure.
TFLM addresses the resource constraints faced with running DL models on embedded systems with 32-bit MCUs and a few 100kB of memory, without hardware-specific customizations \cite{david2021tensorflow}. 
In TFLM, the trained model is interpreted by the application code by giving it pre-processed input data and then performing post-processing of model output and output handling \cite{warden_situnayake_2020}.

A large part of modern CV research and development has been driven by approaches unconstrained by hardware infrastructure, resulting in very large DL models. 
To democratize CV on the edge \cite{david2021tensorflow}, it is necessary to take a hardware-aware approach to ML engineering, so that inference can be performed on lean IoT systems while maintaining compatibility to multiple MCU platforms.

Lack of SE expertise in both IoT embedded vision systems and ML engineering creates an open gap in SE approaches required for TinyML powered IoT embedded vision applications.
A summary of key challenges involved in TinyML engineering is listed in Table \ref{tab:challengestinyml}. 
In addition, the challenges are grouped under key milestones in the TinyML engineering lifecycle.

\section{Research Methodology}
\label{sec:Methodology}
\begin {table*}[htbp]
    \caption{Protocol summary of systematic literature review}
    \begin{tabular}{p{0.2\linewidth} | p{0.75\linewidth}}
        \hline
        Research Questions & RQ1: Which widely used SE practices from large production AI/ML engineering are applicable to TinyML engineering? \\
         & RQ2: Which widely used SE practices from traditional embedded systems engineering are applicable to TinyML engineering? \\\hline
        
        Sample Search strings & "TinyML" AND "Computer Vision" \\
         & "Software Engineering" AND "Edge Computer Vision" \\
         & "Software Engineering" AND "Machine Learning" \\
         & "Software Engineering" AND "Embedded Systems" \\\hline
         
        Search strategy & DB search: arXiv, IEEE, ACM Digital Library, Springer Link \\
         & Backward and forward snowballing using Google Scholar \\
         & Manual search using TinyML.org and Google Search \\\hline
        
        Inclusion Criteria & The paper is written in English \\
         & Grey-literature is accepted since TinyML is quite recent \\\hline
         
        Exclusion Criteria & The paper is related to software engineering applied to traditional computing \\
         & TinyML powered by non-TFLM (TensorFlow Lite Micro) frameworks \\\hline
         
        Study type & Primary and Secondary studies \\\hline
    \end{tabular}
    \label{tab:protocol}
\end{table*}

The previous section described the numerous challenges faced by TinyML application developers. Our systematic literature survey addresses two research questions to summarize SE best practices reported in the literature and visualizes a streamlined workflow for TinyML engineering. TinyML launched only in 2018 and hence, there is a lack of published research into software engineering approaches and best practices. Our research performed a systematic literature review of related prior art in TinyML, Machine Learning, Computer Vision, and Embedded Systems, with the overall goals described here:
\begin{itemize}
    \item Summarization of SE best practices applicable to TinyML engineering for computer vision applications.
    \item Visualization of a SE workflow that can enable quick and easy-to-manage TinyML engineering for developing and deploying CV applications on edge.
    \item Suggestion of ideas for new developer support features in production tools (IDE, packages, test tools, etc.) to accelerate market adoption of TinyML for CV applications.
\end{itemize}

\subsection{Research Questions (RQs)}
We identified two Research Questions (RQs) that span SE challenges and approaches in large ML domain as well as IoT.

\begin{itemize}
\item \textbf{RQ1: Which widely used SE practices from large production AI/ML engineering are applicable to TinyML engineering?}
    \begin{itemize}
        \item Need: To identify SE practices used in generic large AI/ML systems as well as CV-centric large AI/ML systems from a thorough literature survey of state-of-art while determining their applicability to TinyML engineering.
        \item RQ1 aggregates SE solutions already prevalent in practice while ensuring their applicability to challenges imposed on the model size by TinyML constraints.
    \end{itemize}
\item \textbf{RQ2: Which widely used SE practices from traditional embedded systems engineering are applicable to TinyML engineering?}
    \begin{itemize}
        \item Need: To identify SE practices used in embedded systems engineering from a thorough literature survey of state-of-art while determining their applicability to TinyML engineering.
        \item RQ2 aggregates SE solutions that are already prevalent in practice while ensuring their applicability to challenges imposed by TinyML vision applications.
    \end{itemize}
\end{itemize}

\subsection{Research protocol}
TinyML and TFLM were introduced in 2018, and the published scientific literature is limited. Only a handful of papers pertain to embedded vision applications. Additionally, the TinyML literature is mostly gray literature, tutorials, workshops, talks, and blogs. To address challenges in TinyML engineering reported in the literature, the systematic literature survey focused on aggregating solutions from a broad spectrum of information sources.

We gathered 43 information sources from arXiv, IEEE, ACM Digital Library, Springer Link, ML conference tutorials, and workshops using sample search queries described in Table \ref{tab:protocol}. Then, we studied the abstract and conclusions from the gathered literature to refine the source pool to 20. Finally, we forward and backward snowballed on Google Scholar, Google Search, and related Blogs to derive 13 additional sources. In total, we have 33 information sources as the primary pool for this study.


\section{Results and Discussion}
\label{sec:Results}
The first section ties various findings into the research questions defined at the start of the project. 
The second section ties the SE approaches and best practices to derive a SE workflow for TinyML CV application developers. 
In the interest of brevity, only the most compelling results relevant to IoT embedded vision systems are presented.
Challenges and solutions in TinyML engineering are presented in Table \ref{tab:resulttable1} and \ref{tab:resulttable2}, with classification based on steps involved in TinyML workflow process.
Answers from RQ1 and RQ2 were evaluated for applicability to TinyML engineering of edge CV applications through thorough study.

\begin{table*}[htbp]
    \centering
    \caption{SE challenges and solutions in TinyML engineering for edge CV apps}
    \label{tab:resulttable1}
    \begin{tabular}{p{0.15\linewidth} | p{0.35\linewidth} | p{0.35\linewidth} | p{0.06\linewidth}}
        \hline
        WORKFLOW STEP & CHALLENGE & PROPOSED SOLUTION(S) & ORIGIN\\\hline
        
        Business (Customer) Understanding & TinyML engineering requires expertise in AI, ML, CV, and Embedded Systems. & Decoupling teams which create systems, data, and model data is helpful to carry out work concurrently \cite{8804457}. & RQ1 \\
        
         & Customer needs are changing dynamically. & In order to streamline the TinyML engineering process, customer needs must be version-controlled to pursue incremental improvement on tail features requested by the customer \cite{giray2021software}. & RQ1 \\
        
        \hline
        
        System Requirements & Deployment in real-world environments is challenging. & Define common implicit assumptions related to visual data (illumination, shadows, etc). & RQ1 \\
         
         & Mismatch in hardware systems and portability of models across a family of hardware is required. & Need to manage the combinatorial explosion of hardware devices using strict control of device types and chipsets used in the TinyML systems~\cite{se4cvsystems}. & RQ2 \\
         
         & Performance degradation in a real-world scenario is a major challenge to TinyML CV systems on the edge. & Design decisions must be driven by environmental challenges that the TinyML CV application is likely to experience. Data augmentation of the training data set is necessary \cite{se4cvsystems}. & RQ1 \\
         
        \hline
        
        Model Requirements & Determination of models applicable to TinyML projects is not well documented. & TinyML systems need to select target MCU hardware for their projects based on requirements of model size, inference speed, and power limitations necessary for the application \cite{banbury2021benchmarking}. & RQ2 \\
        
         & Determining useful models for lean edge is not an easy task. & Constrained Neural Architectural Search (NAS) can be used to narrow down model architectures relevant to the system design choices \cite{fedorov2019sparse} \cite{paissan2021phinets}. & RQ1 \\
        
        \hline
        
        Data Requirements & Data sets available can have very limited data. & Test-driven development can be used to ensure test data is not used during training as well as to identify ways to test corner cases based on system specs \cite{giray2021software}. & RQ2 \\
        
        \hline
        
        System Design & Model development tools are tedious to maintain for different target hardware systems. & Setting up TinyML TFLM framework infrastructure correctly at start will help iterate faster \cite{david2021tensorflow}. & RQ1 \\
        
         & Model development involved high complexity when catering to an array of applications. & Model management and version control are very important for streamlining TinyML development efforts \cite{se4cvsystems}. & RQ1 \\
        
         & Vision data related to videos often contain redundant information temporally, and it leads to additional unnecessary computation on the lean edge hardware. & Event-based approaches to visual information processing can be applied to reduce the burden imposed by redundant computations \cite{amossironi}. & RQ1 \\
        
        \hline
        
        Data Engineering & Data derived from different sensors and sources can have various file formats. & No direct access to data, only programmatic interfaces (8-bit RGB image instead of image.jpg)~\cite{se4cvsystems}. & RQ1 \\
        
         & Real-world conditions might not be represented in distributions found on data sets. & Data augmentation can be used to simulate real-world conditions (scaling, noise, shadows, color, lighting conditions, etc.). & RQ1 \\
        
        \hline
        
        Model Training & Hyperparameter search process can experience issues in reproducibility. & Usage of deterministic randomness (using exposed random seeds) is helpful for iteration with hyperparameter tuning \cite{se4cvsystems}. & RQ1 \\
        
         & Training TinyML CV models with a performance driven perspective is hard. & Building training pipeline by adding verification-based counter-examples will be helpful \cite{se4cvsystems} \cite{alessio}. & RQ1 \\
        
         & TinyML application developers might lack deep expertise in ML and DL. & TinyML application developers must leverage AutoML tools to automate the training process, especially when a large amount of training and validation data is available. & RQ1 \\

    \end{tabular}
\end{table*}

\begin{table*}[htbp]
    \centering
    \caption{SE challenges and solutions in TinyML engineering (continued)}
    \label{tab:resulttable2}
    \begin{tabular}{p{0.15\linewidth} | p{0.35\linewidth} | p{0.35\linewidth} | p{0.06\linewidth}}
        \hline
        WORKFLOW STEP & CHALLENGE & PROPOSED SOLUTION(S) & ORIGIN\\\hline
        
        Model Training (contd.) & Web automation of TinyML engineering. & EdgeImpulse and Qeexo among other startups offer AutoML capabilities that are delivered over web, directly into edge CV applications \cite{armblueprint_2021}. & RQ1 \\
        
        \hline
        
        Model Evaluation & TinyML models need to be robust to take care of environmental anomalies and noise. & Test-driven development must rely on measuring local and global adversarial robustness \cite{alessio}. & RQ1 \\
        
         & Lack of standardized TinyML performance metrics in reported literature. & MLPerfTiny benchmark has been published by MLCommons to provide the first industry standard benchmark suite for ultra low power ML systems \cite{banbury2021mlperf}. & RQ1 \\
        
        \hline
        
        Model Compilation & Performance cannot be measured until an application is deployed on the edge device system. & Compilers must be capable of providing access to performance metrics for MCUs as they are typically single-threaded \cite{david2021tensorflow}. & RQ1 \\
        
         & Compiler settings need to be optimized for hardware system targeted by the system design choices. & Hardware aware compiler is helpful to take desirable tradeoffs automatically without the need for understanding of MCU architectures \cite{lai2018cmsisnn} \cite{xu2020edge}. & RQ1 \\
        
        \hline
        
        Compiled Model Evaluation & It is hard to debug compiled model performance on the edge device due to compute constraints. & Device simulators and emulators must be used in TinyML development. Device simulators can help contrast performance of raw models and various flavors of compiled TinyML models \cite{se4cvsystems}. & RQ2 \\
        
        \hline
        
        Application Design & Design tradeoffs are hard to optimize until the model is optimized. & Automated tools can aid the search for tradeoffs between performance, accuracy, and power \cite{banbury2021micronets}. & RQ1 \\
        
        \hline
        
        Application Evaluation & Application issues are harder to debug since the system involves heuristic code, as well as trained TinyML model. & Review of bug taxonomy for IoT applications, can help avoid critical issues in application code~\cite{Makhshari2021IoTBA}~\cite{islam2019comprehensive}. & RQ2 \\
        
         & Testing TinyML models can be derailed due to bias in tests. & Dedicated QA teams can help with fair debug with developer bias \cite{se4cvsystems}. & RQ2 \\
        
        \hline
        
        Application Deployment & Real-world deployment of TinyML is hard even after successful validation of models in training environment. & Reproducibility is driven when algorithms, parameters, and data/labels are ‘uniquely identifiable’ and ‘retrievable’ in real-world. \cite{se4cvsystems}. & RQ1 \\
        
         & Fixing bugs in embedded application deployment is hard. & Over-the-air (OTA) updates must be available to seamlessly deliver bug fixes and upgrades for both firmware and TinyML model \cite{mehta_2021}. & RQ2 \\
         
        \hline
        
        Application Monitoring & Performance improvements in the field is challenging due to limitations on edge hardware. & Continuous Integration (CI) of Labeling and Re-training using feedback from monitoring can help drive application performance with new data distributions are encountered in the field \cite{se4cvsystems}. & RQ1 \\
        
         & It is hard to measure the performance of the TinyML CV application in the field. & Device-level monitoring tools can be used to improve debuggability in the field \cite{se4cvsystems}. & RQ2 \\
        
        \hline

    \end{tabular}
\end{table*}

\begin{figure*}[htbp]
    \centering
    \includegraphics[width=\linewidth]{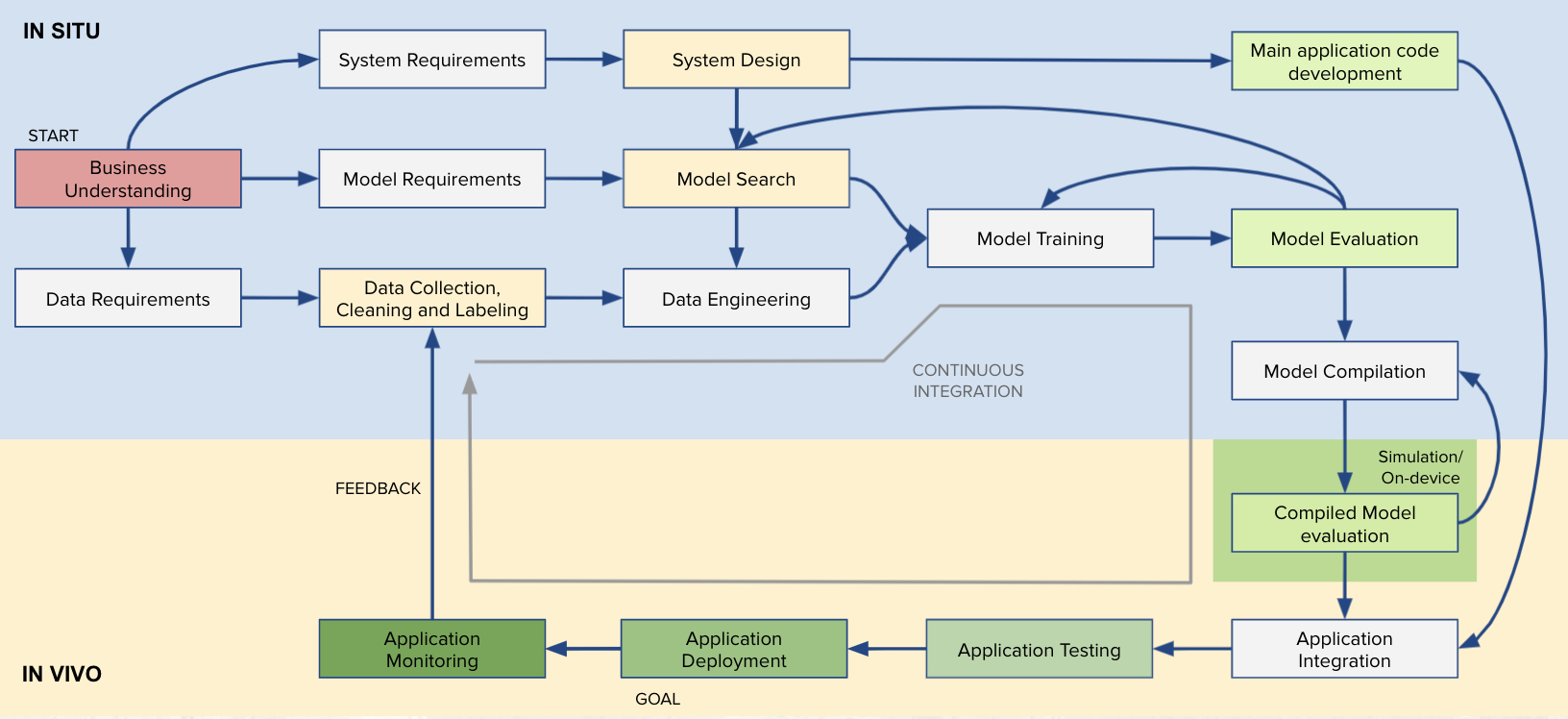}
    \caption{SE workflow for TinyML engineering for CV applications}
    \label{fig:seworkflow}
\end{figure*}

\subsection{RQ1 findings}
Established SE practices in large-scale AI/ML and CV engineering are applicable broadly to solve many of the challenges in IoT embedded vision engineering. 
However, there is a significant need for hardware-aware Neural Architecture Search (NAS) approaches, hardware-aware co-optimization approaches, reference datasets for IoT embedded vision applications, and standardized benchmarking techniques and metrics.

\subsection{RQ2 findings}
It is to be noted that established SE practices from embedded/IoT engineering largely apply to IoT embedded vision systems. 
However, there is a significant need for portable libraries for pre-processing and post-processing functions that are key components of the ML inference architecture. 
SE practices need tools capable of design space search and automatic version control for deployment-ready TinyML models.

\subsection{SE workflow for TinyML engineering}

Based on the results of the systematic literature survey, we see the need to define a software engineering workflow specifically to cater to TinyML engineering. 
Inspired by CRISP-DM, Microsoft ML workflow, and TFLM workflows, we present a modified workflow in Figure \ref{fig:seworkflow}. The workflow can be adopted by CV application developers alongside results summarized by this systematic literature survey.
The proposed SE workflow divides the engineering process into steps that are suggested to be performed `in situ' aka `training environment with CPU and GPU' and to be performed 'in vivo' aka 'lean edge hardware performing inference in real-world deployment'.

It is beneficial to break down the business requirements (customer requirements) into three inter-related components: model requirements, data requirements, and system requirements. 
This approach helps delegate ownership of delivering individual components to different individuals or teams. 
It also enables autonomy for those teams to understand requirements to apply their expertise in system design, model search, and setting up pipelines for data collection, cleaning, and labeling. 
Once the system design is generated, it will provide additional inputs to constrain model search and data engineering. 
Given that the application must run on a lean MCU resource, system design provides the boundary conditions on computing-memory complexity of algorithms used in data engineering (pre-processing and post-processing) and machine learning.

Main application development can be carried out concurrently with model development. 
Concurrent development allows embedded developers to develop and debug heuristic pipelines to perform sensing and actuation based on expected outcomes from the ML model. 
They can also debug this code using calibration functions to emulate outputs of the final ML model.
The model development process can continue after model search and data engineering steps. 
Using best practices in model training, a suitable model can be derived and evaluated against a validation data set. 
Feedback from the evaluation step can lead model developers into model search or model training to reconfigure the model or its hyper-parameters. 

Once the model validation is satisfactory, the model can proceed to the compilation steps, which are automatically taken care of by the TFLM framework. 
Hardware-specific libraries can be used during this process to take advantage of custom hardware features available in target MCUs.
After the model is compiled successfully, compiled model evaluation can be carried out in a simulator in the training environment or on-device inference environment based on the availability. 
The advantage of executing this step in a simulator environment has been described earlier in the results section. 
If there is a significant drop in model performance after compilation, the feedback must be relayed back to the model compilation step so that necessary actions can be taken.

Once the compiled model delivers performance comparable to the original ML model, the model can be delivered to the application integration team, which will integrate the model into the application code, followed by application testing with test data set and deployment in the field. 
Monitoring application performance in the real-world situations can provide valuable feedback, which can be used for Continuous Integration (CI) and guide the augmentation or addition of suitable new data from real-world distributions.

\subsection{Future IoT embedded vision developer support tools}

We are proposing certain key features for next-generation IoT embedded vision developer support tools.
These features enable a better developer experience, reduce bugs, improve debuggability, and productize scale IoT embedded vision applications.
\begin{enumerate}
    \item Embedded IoT developer tools must integrate TinyML frameworks such as TFLM, through an interactive user-interface.
    \item SE workflow must be integrated into the toolchain to guide developers through the application design, development, and deployment process.
    \item Allow developers to import MCU models into the tool and visualize the MCU architecture. Allow comparison of different MCU choices to guide design choices based on application performance targets (FPS, TOPS, Precision support, etc).
    \item Allow emulation of devices to test the compatibility of compiled models prior to deployment on physical hardware. This creates an opportunity to visualize application and model operation without significant constraints on measurability.
    \item Automatically capture device capabilities and set constraints into the design space to guide the TinyML development process.
    \item Allow integration of custom libraries into the toolchain to extend generic compiler capabilities to help developers leverage unique hardware features on the target MCU.
    \item Support multiple device management for IoT embedded systems including the capability to send firmware, code, and model updates over-the-air (OTA).
    \item Allow configuration and interaction with a large number of IoT embedded vision devices through RESTful API interfaces.
\end{enumerate}


\section{Threats to Validity}
\label{sec:Threats}
As TinyML is evolving at a fast pace, framework designs are bound to evolve rapidly and new SE approaches might be necessary. 
Hence, the study lacks the perspective that may be obtained through surveys and interviews of TinyML CV application developers in the industry.
Amount of TinyML literature is limited since TinyML and TFLM concepts were introduced very recently (2018). 
It is hard to quantify existing TinyML SE practices from only a handful of peer-review publications.

As the field evolves rapidly, additional challenges and best practices will come to light which can challenge some of the conclusions drawn by our study.
Novel model architectures, model search approaches, TinyML development techniques may significantly alter the proposed SE workflow and developer tool features.

\section{Conclusion}
\label{sec:Conclusion}
There is significant scope for more profound research into SE approaches for production-scale engineering of IoT embedded vision applications. 
Hardware-aware generic TinyML compiler tool features will be required to extract full performance out of lean edge devices. 
At the same time, application portability continues to challenge fixed-form compilers that focus on narrow optimizations for highly specific hardware systems.

The proposed SE workflow is useful for concurrent work by model developers, application code developers, and data engineers, with high potential to evolve with TinyML research. 
This workflow also accelerates the TinyML engineering process by delegating work to teams with the necessary expertise. 
The integration of the CI approach into the workflow provides a robust approach for refining the TinyML product in the field without requiring expensive recalls or repairs.

Further experimental research on the proposed SE best practices and workflow will help validate and refine the SE approaches for production-scale engineering TinyML CV applications.

\balance{}

\bibliographystyle{abbrv}
\bibliography{sigproc}

\end{document}